\documentclass[twocolumn]{aastex63}
\usepackage{url}
\usepackage{threeparttable}
\usepackage{ulem}

\usepackage{natbib}
\usepackage{appendix}
\usepackage{amsmath}
\usepackage{multirow}

\def\tcra{\textcolor{black}}

\def\oii{O\,\textsc{ii}}
\def\oiii{O\,\textsc{iii}}
\def\sii{S\,\textsc{ii}}
\def\ciii{C\,\textsc{iii}}

\newcommand{\approptoinn}[2]{\mathrel{\vcenter{
  \offinterlineskip\halign{\hfil$##$\cr
    #1\propto\cr\noalign{\kern2pt}#1\sim\cr\noalign{\kern-2pt}}}}}

\newcommand{\appropto}{\mathpalette\approptoinn\relax}

\received{Jan 17, 2023}
\revised{Aug 11, 2023}
\accepted{Aug 22, 2023}
\shorttitle{
Redshift Evolution of the Electron Density at $z\sim 0-9$
}

\shortauthors{Isobe et al.}
\graphicspath{{./}{figures/}}

\begin{document}

\title{
Redshift Evolution of the Electron Density in the ISM at $z\sim 0-9$\\ 
Uncovered with JWST/NIRSpec Spectra and Line-Spread Function Determinations
}

\author[0000-0001-7730-8634]{Yuki Isobe}
\affiliation{Institute for Cosmic Ray Research, The University of Tokyo, 5-1-5 Kashiwanoha, Kashiwa, Chiba 277-8582, Japan}
\affiliation{Department of Physics, Graduate School of Science, The University of Tokyo, 7-3-1 Hongo, Bunkyo, Tokyo 113-0033, Japan}

\author[0000-0002-1049-6658]{Masami Ouchi}
\affiliation{National Astronomical Observatory of Japan, 2-21-1 Osawa, Mitaka, Tokyo 181-8588, Japan}
\affiliation{Institute for Cosmic Ray Research, The University of Tokyo, 5-1-5 Kashiwanoha, Kashiwa, Chiba 277-8582, Japan}
\affiliation{Kavli Institute for the Physics and Mathematics of the Universe (WPI), University of Tokyo, Kashiwa, Chiba 277-8583, Japan}

\author[0000-0003-2965-5070]{Kimihiko Nakajima}
\affiliation{National Astronomical Observatory of Japan, 2-21-1 Osawa, Mitaka, Tokyo 181-8588, Japan}

\author[0000-0002-6047-430X]{Yuichi Harikane} 
\affiliation{Institute for Cosmic Ray Research, The University of Tokyo, 5-1-5 Kashiwanoha, Kashiwa, Chiba 277-8582, Japan}

\author[0000-0001-9011-7605]{Yoshiaki Ono}
\affiliation{Institute for Cosmic Ray Research, The University of Tokyo, 5-1-5 Kashiwanoha, Kashiwa, Chiba 277-8582, Japan}

\author[0000-0002-5768-8235]{Yi Xu}
\affiliation{Institute for Cosmic Ray Research, The University of Tokyo, 5-1-5 Kashiwanoha, Kashiwa, Chiba 277-8582, Japan}
\affiliation{Department of Astronomy, Graduate School of Science, The University of Tokyo, 7-3-1 Hongo, Bunkyo, Tokyo 113-0033, Japan}

\author{Yechi Zhang}
\affiliation{Institute for Cosmic Ray Research, The University of Tokyo, 5-1-5 Kashiwanoha, Kashiwa, Chiba 277-8582, Japan}
\affiliation{Department of Physics, Graduate School of Science, The University of Tokyo, 7-3-1 Hongo, Bunkyo, Tokyo 113-0033, Japan}

\author{Hiroya Umeda}
\affiliation{Institute for Cosmic Ray Research, The University of Tokyo, 5-1-5 Kashiwanoha, Kashiwa, Chiba 277-8582, Japan}
\affiliation{Department of Physics, Graduate School of Science, The University of Tokyo, 7-3-1 Hongo, Bunkyo, Tokyo 113-0033, Japan}




\begin{abstract}
We present electron densities $n_{\rm e}$ in the inter-stellar medium (ISM) of star-forming galaxies at $z=4-9$ observed by the JWST/NIRSpec GLASS, ERO, and CEERS programs. We carefully evaluate line-spread functions of the NIRSpec instrument as a function of wavelength with the calibration data of a planetary nebula taken onboard, and obtain secure [\oii]$\lambda\lambda$3726,3729 doublet fluxes for 14 galaxies at $z=4.02-8.68$ falling on the star-formation main sequence with the NIRSpec high and medium resolution spectra. We thus derive the electron densities of singly-ionized oxygen nebulae with the standard $n_{\rm e}$ indicator of [\oii] doublet, and find that the electron densities of the $z=4-9$ galaxies are $n_{\rm e}\gtrsim 300$ cm$^{-3}$ significantly higher than those of low-$z$ galaxies at a given stellar mass, star-formation rate (SFR), and specific SFR. Interestingly, typical electron densities of singly ionized nebulae increase from $z=0$ to $z=1-3$ and $z=4-9$, which is approximated by the evolutionary relation of $n_{\rm e}\propto(1+z)^{p}$ with $p\sim 1-2$.
Although it is not obvious that the ISM property of $n_{\rm e}$ is influenced by global galaxy properties, these results may suggest that nebula densities of high-$z$ galaxies are generally high due to the compact morphologies of high-$z$ galaxies evolving by $r_{\rm e} \appropto (1+z)^{-1}$ ($r_{\rm vir} \propto (1+z)^{-1}$) for a given stellar (halo) mass whose inverse square corresponds to the $p\sim 2$ evolutionary relation. 
The $p\sim 1-2$ evolutionary relation can be explained by a combination of the compact morphology and the reduction of $n_{\rm e}$ due to the high electron temperature of the high-$z$ metal poor nebulae.
\end{abstract}

\keywords{Galaxy formation (595); Galaxy structure (622); Star formation (1569); Dwarf galaxies (416)}


\section{Introduction} \label{sec:intro}
\begin{figure*}[t]
    \centering
    \includegraphics[width=18.0cm]{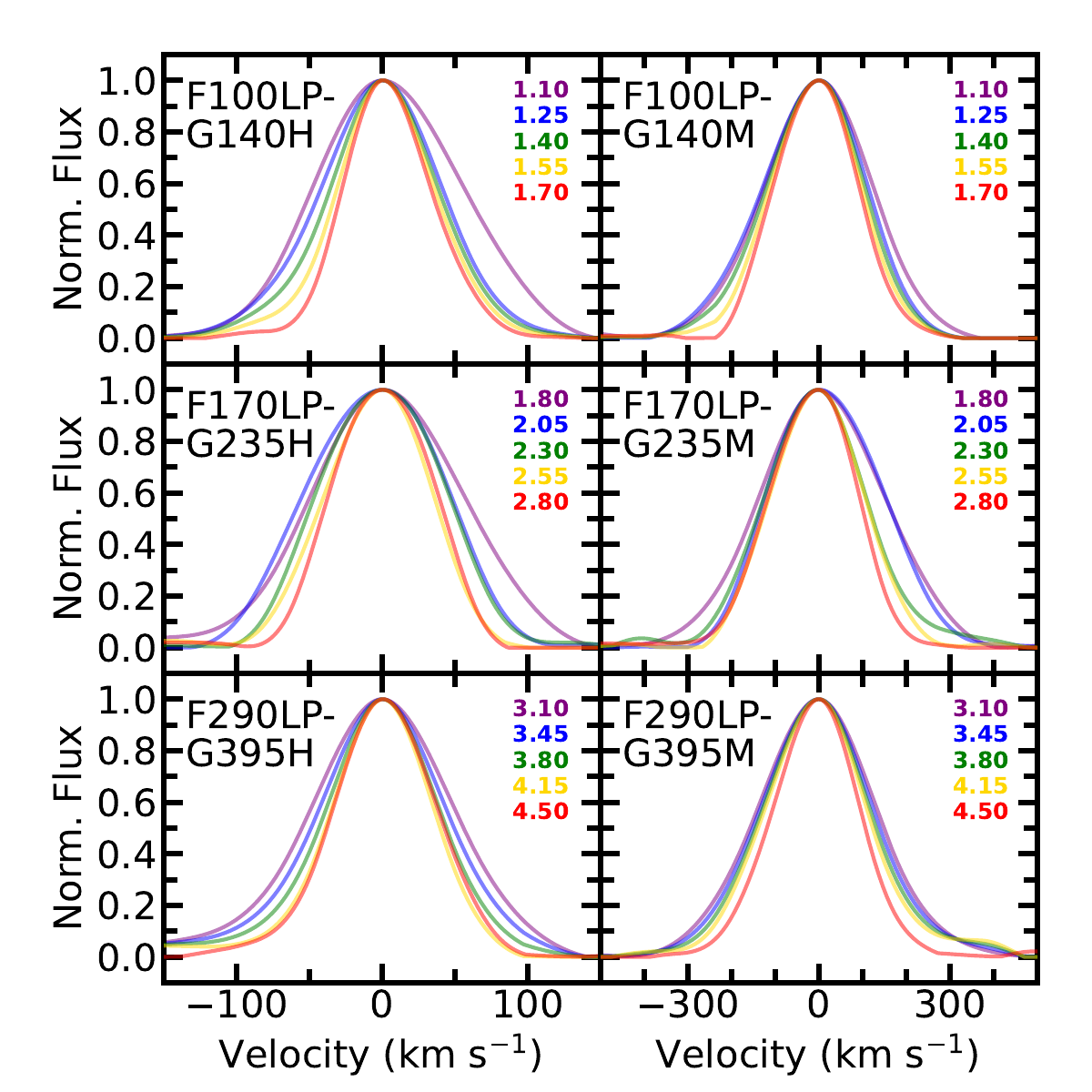}
    \caption{LSF as a function of velocity. In each panel, the purple, blue, green, yellow, and red curves represent LSFs obtained at the wavelengths whose values correspond to the numbers with the same color listed at the top right corner (unit: $\mu$m).}
    \label{fig:lsf}
\end{figure*}

\begin{figure*}[t]
    \centering
    \includegraphics[width=18.0cm]{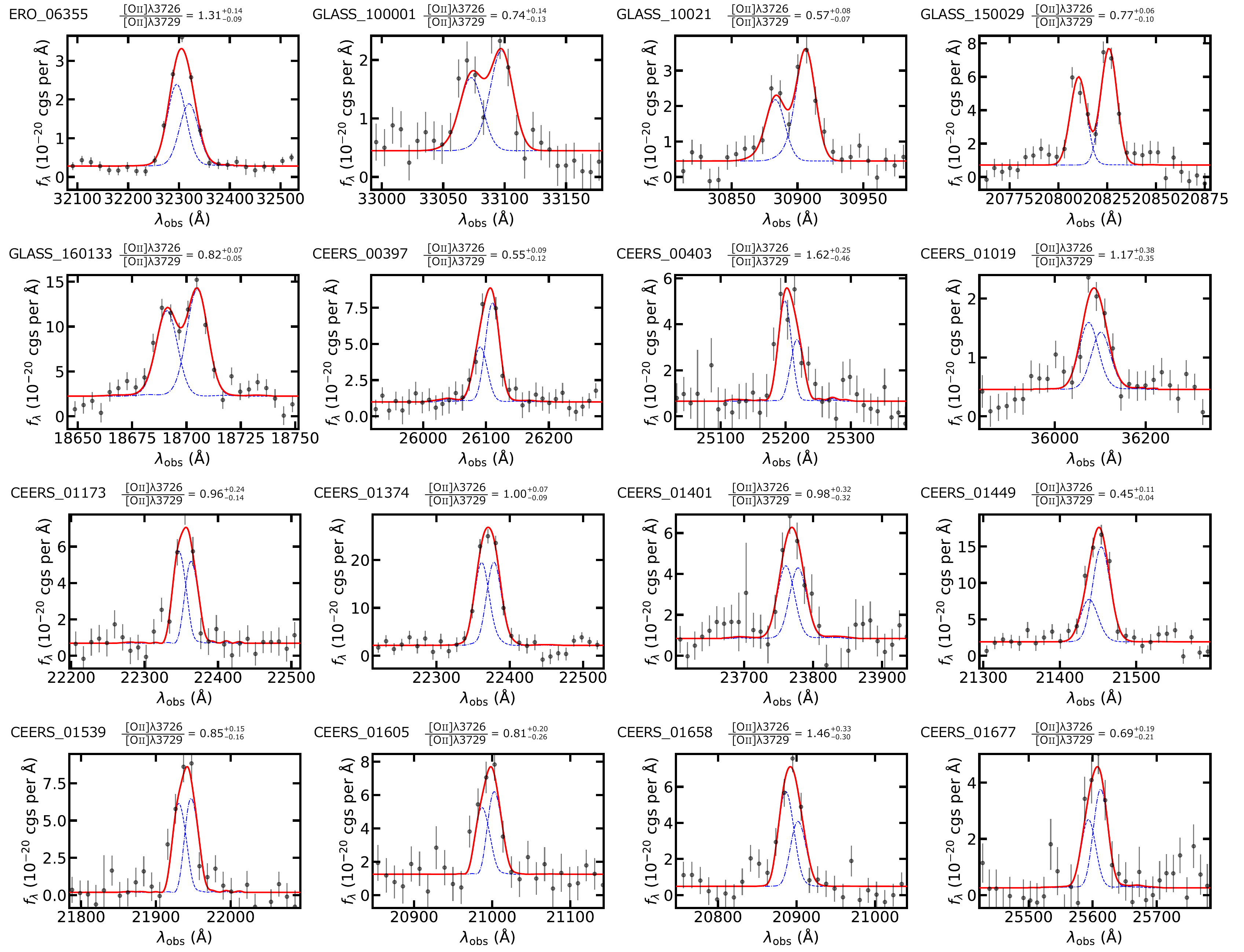}
    \caption{\tcra{Our fitting results} of the LSF-convolved Gaussian functions to the [\oii] doublet. The gray dots denote data points with spectral errors. The red solid curve represents a total profile of the [\oii] doublet. The blue dashed curve and the blue dashdotted curve show profiles of [\oii]$\lambda3726$ and [\oii]$\lambda3729$, respectively.}
    \label{fig:o2}
\end{figure*}

Properties of interstellar medium (ISM) have spectroscopically investigated \tcra{by many studies using} photoionization models \citep[e.g.,][]{Nagao2012,Nakajima2014,Feltre2016,Steidel2016,Sanders2016,Schaerer2018,Kewley2019,Harikane2020,Berg2021,Umeda2022,Matsumoto2022}.
Particularly, electron densities $n_{\rm e}$ of H\,\textsc{ii} regions are one of the key quantities to describe physical states of ISM.
The electron density is usually derived from density-sensitive emission-line ratios such as [\oii]$\lambda$3726/[\oii]$\lambda$3729, [\sii]$\lambda$6716/[\sii]$\lambda$6731, and [\ciii]$\lambda$1907/[\ciii]$\lambda$1909 \citep[e.g.,][]{Kewley2019}.
Large spectroscopic surveys have found that star-forming galaxies with $M_{*}\sim10^{10}\ M_{\odot}$ have $n_{\rm e}$ values of low-ionization regions (\oii\ and \sii) increasing from $n_{\rm e}\sim30$ to $\sim200$ cm$^{-3}$ with the redshift in the range of $z\sim0$--3 \citep[e.g., ][]{Steidel2014,Shimakawa2015,Sanders2016,Kashino2017,Davies2021}.
Beyond $z\gtrsim4$, $n_{\rm e}$ values had not been investigated due to the lack of high-sensitivity near-infrared spectrographs before the arrival of \textit{James Webb Space Telescope} (JWST), except for a few suggestions based on doubly-ionized regions such as $n_{\rm e}$(\ciii$)\sim4$--$13\times10^{4}$ cm$^{-3}$ at $z=10.957$ \citep{Jiang2021} and $n_{\rm e}$(\oiii$)\lesssim500$ cm$^{-3}$ at $z=7.133$ \citep{Killi2023}.
Recently, using the flux ratio of [\oiii]$\lambda$5007 from the JWST/NIRSpec Early Release Observations (ERO; \citealt{Pontoppidan2022}) and [\oiii]$\lambda$88$\mu$m from ALMA observations, \citet{Fujimoto2022arXiv} have reported that a $z=8.496$ dwarf galaxy, s04590, has a $n_{\rm e}$(\oiii) value of $220^{+170}_{-100}$ cm$^{-3}$.

Although the $n_{\rm e}$(\oiii\ or \ciii) values are likely to be higher than those of $z\sim0$ star-forming galaxies (e.g., $n_{\rm e}$(\oii$)\sim30$ cm$^{-3}$; \citealt{Davies2021}), we need to discuss the redshift evolution of $n_{\rm e}$ at similar ionization regions (i.e., \oii\ and \sii, rather than \oiii\ or \ciii).
We should also check if $n_{\rm e}$ values change with other galaxy properties such as $M_{*}$ or star-formation rate (SFR).
In this paper, we aim to measure $n_{\rm e}$(\oii) values of high-$z$ ($z>4$) star-forming galaxies identified by NIRSpec public surveys.
Comparing the $n_{\rm e}$ values with those of lower-$z$ star-forming galaxies, we also investigate the redshift evolution of $n_{\rm e}$ and the potential dependences on $M_{*}$ and SFR.

This paper is organized as follows.
Section \ref{sec:datsam} explains our data and sample.
Our analysis is described in Section \ref{sec:analysis}.
We report our results and discuss them in Section \ref{sec:res}.
Our findings are summarized in Section \ref{sec:sum}.
We assume a standard $\Lambda$CDM cosmology with parameters of ($\Omega_{\rm m}$, $\Omega_{\rm \Lambda}$, $H_{0}$) = (0.3, 0.7, 70 km ${\rm s}^{-1}$ ${\rm Mpc}^{-1}$). 

\section{Data and Sample} \label{sec:datsam}
We use JWST/NIRSpec data of the Early Release Observations (ERO; \citealt{Pontoppidan2022}) taken in the SMACS 0723 lensing cluster field (hereafter ERO data), the GLASS \citep{Treu2022} survey (hereafter GLASS data), and the CEERS \citep{Finkelstein2023} survey (hereafter CEERS data).
The ERO data were taken with medium resolution ($R\sim 1000$) filter-grating pairs of F170LP-G235M and F290LP-G395M covering the wavelength ranges of $1.7$--$3.1$ and $2.9$--$5.1\ \mu$m, respectively.
The total exposure time of the ERO data is 4.86 hr for each filter-grating pair.
The GLASS data were taken with high resolution ($R\sim 2700$) filter-grating pairs of F100LP-G140H, F170LP-G235H, and F290LP-G395H covering the wavelength ranges of $1.0$--$1.6$, $1.7$--$3.1$ and $2.9$--$5.1\ \mu$m, respectively.
The total exposure time of the GLASS data is 4.9 hr for each filter-grating pair.
The CEERS data were taken with medium resolution filter-grating pairs of F100LP-G140M, F170LP-G235M, and F290LP-G395M covering the wavelength ranges of $1.0$--$1.6$, $1.7$--$3.1$ and $2.9$--$5.1\ \mu$m, respectively.
The total exposure time of the CEERS data is 0.86 hr for each filter-grating pair.

We use spectroscopic data reduced by \citet{Nakajima2023arXiv}.
\citet{Nakajima2023arXiv} take the raw data from the MAST archive and conduct level-2 and 3 calibrations, using the JWST Science Calibration Pipeline with the reference file of jwst\_1028.pmap whose flux calibration is based on in-flight flat data.
\tcra{In addition to the read-out noise and Poisson noise outputted by the JWST/NIRSpec pipeline, \citet{Nakajima2023arXiv} have added standard deviation of the residual background to evaluate the uncertainty of background subtraction.}
Checking the data, we identify 5, 14, and 55 galaxies at $z>4$ in the ERO, GLASS, and CEERS data, respectively\tcra{, whose redshifts by simultaneous fitting of the H$\beta$ and [\oiii]$\lambda\lambda4959,5007$.}
We omit galaxies with S/N ratios of the [\oii] doublet less than 10 so that we can obtain galaxies with well-determined $n_{\rm e}$.
After the selection, we obtain 1, 4, and 11 galaxies from the ERO, GLASS, and CEERS data, respectively.
Hereafter we refer to the 16 ($=1+4+11$) galaxies as JWST galaxies.

\section{Analysis} \label{sec:analysis}
\begin{table*}[t]
    \begin{center}
    \caption{Electron density and other properties}
    \label{tab:ne}
    \begin{tabular}{ccccccccc} \hline \hline
        Name & R.A. & Decl. & $z$ & $n_{\rm e}$(\oii) & 
        \tcra{$\frac{[\rm \oii]\lambda3726}{[\rm \oii]\lambda3729}$} & $\sigma_{\rm int}$ & $\log(M_{*})$ & $\log(\rm SFR)$ \\
        & deg & deg & & cm$^{-3}$ & & km s$^{-1}$ & $M_{\odot}$ & $M_{\odot}$ yr$^{-1}$ \\
		(1) & (2) & (3) & (4) & (5) & (6) & (7) & (8) & (9) \\ \hline
        ERO\_06355 & 110.8267392 & $-73.4537444$ & 7.6651 & $1433^{+752}_{-335}$ & $1.31^{+0.14}_{-0.09}$ & 113 & $8.77^{+0.08\,{\rm a}}_{-0.01}$ & $1.41^{+0.01\,{\rm a}}_{-0.01}$ \\
        GLASS\_100001 & 3.6038450 & $-30.3822350$ & 7.8737 & $118^{+217}_{-118}$ & $0.74^{+0.14}_{-0.13}$ & 69 & $9.15^{+0.55\,{\rm a}}_{-0.00}$ & $0.95^{+0.10\,{\rm a}}_{-0.13}$ \\
        GLASS\_150029 & 3.5771664 & $-30.4225760$ & 4.5837 & $174^{+137}_{-174}$ & $0.77^{+0.06}_{-0.10}$ & 38 & $9.12^{+0.03\,{\rm a}}_{-0.33}$ & $1.04^{+0.02\,{\rm a}}_{-0.02}$ \\
        GLASS\_160133 & 3.5802754 & $-30.4244040$ & 4.0150 & $252^{+161}_{-103}$ & $0.82^{+0.07}_{-0.05}$ & 59 & $8.11^{+0.41\,{\rm a}}_{-0.22}$ & $1.16^{+0.01\,{\rm a}}_{-0.01}$ \\
        CEERS\_00397 & 214.8361970 & $52.8826930$ & 6.0005 & $0^{+1}_{-0}$ & $0.55^{+0.09}_{-0.12}$ & 81 & $8.45^{+0.53\,{\rm a}}_{-0.06}$ & $1.76^{+0.02\,{\rm a}}_{-0.03}$ \\
        CEERS\_00403 & 214.8289680 & $52.8757010$ & 5.7609 & $3105^{+3232}_{-2151}$ & $1.62^{+0.25}_{-0.46}$ & 56 & $9.35^{+0.27\,{\rm a}}_{-0.09}$ & $1.20^{+0.06\,{\rm a}}_{-0.07}$ \\
        CEERS\_01019 & 215.0353914 & $52.8906618$ & 8.6791 & $1118^{+3332}_{-969}$ & $1.17^{+0.38}_{-0.35}$ & 123 & $10.12^{+0.12\,{\rm a}}_{-0.11}$ & $1.97^{+0.04\,{\rm a}}_{-0.05}$ \\
        CEERS\_01173 & 215.1542076 & $52.9558470$ & 4.9957 & $574^{+942}_{-331}$ & $0.96^{+0.24}_{-0.14}$ & $\cdots^{\rm c}$ & $9.38^{+1.12\,{\rm b}}_{-1.12}$ & $0.56^{+0.13\,{\rm b}}_{-0.13}$ \\
        CEERS\_01374 & 214.9439110 & $52.8500419$ & 4.9999 & $656^{+254}_{-219}$ & $1.00^{+0.07}_{-0.09}$ & 67 & $9.44^{+1.12\,{\rm b}}_{-1.12}$ & $1.36^{+0.03\,{\rm b}}_{-0.03}$ \\
        CEERS\_01401 & 215.2458005 & $53.0652953$ & 5.3749 & $613^{+1934}_{-613}$ & $0.98^{+0.32}_{-0.32}$ & 112 & $<8.92^{\rm b}$ & $0.86^{+0.09\,{\rm b}}_{-0.09}$ \\
        CEERS\_01539 & 214.9800779 & $52.9426590$ & 4.8841 & $330^{+444}_{-327}$ & $0.85^{+0.15}_{-0.16}$ & $\cdots^{\rm c}$ & $<8.71^{\rm b}$ & $0.80^{+0.05\,{\rm b}}_{-0.05}$ \\
        CEERS\_01605 & 215.0754073 & $52.9975786$ & 4.6309 & $250^{+537}_{-250}$ & $0.81^{+0.20}_{-0.26}$ & 16 & $9.11^{+1.10\,{\rm b}}_{-1.10}$ & $0.53^{+0.06\,{\rm b}}_{-0.06}$ \\
        CEERS\_01658 & 214.9852372 & $52.9242590$ & 4.6038 & $2199^{+5950}_{-1078}$ & $1.46^{+0.33}_{-0.30}$ & 64 & $<8.56^{\rm b}$ & $0.67^{+0.06\,{\rm b}}_{-0.06}$ \\
        CEERS\_01677 & 215.1887384 & $53.0643782$ & 5.8670 & $39^{+470}_{-39}$ & $0.69^{+0.19}_{-0.21}$ & 83 & $9.07^{+0.84\,{\rm b}}_{-0.84}$ & $0.69^{+0.08\,{\rm b}}_{-0.08}$ \\\hline
    \end{tabular}
    \end{center}
    \tablecomments{(1) Name. (2) Right ascension in J2000. (3) Declination in J2000. (4) Redshift. (5) Electron density derived from the [\oii] doublet (Section \ref{subsec:ne}). (6) Velocity dispersion (Section \ref{subsec:sig}). (7) Stellar mass (Section \ref{subsec:prop}). (8) Star-formation rate (Section \ref{subsec:prop}). \\ a: The $M_{*}$ and SFR values are estimated from SED fitting to JWST/NIRCam photometry with prospector \citep{Johnson2021} and H$\beta$ fluxes corrected for slit losses, respectively (\citealt{Nakajima2023arXiv}; Section \ref{subsec:prop}). \\ b: The $M_{*}$ and SFR values are estimated from HST restframe-UV photometry and H$\beta$ fluxes not corrected for slit losses, respectively, because no NIRCam photometry are available for these objects to date (Section \ref{subsec:prop}). \\ c: The $\sigma_{\rm int}$ values are too small to measure (Sections \ref{subsec:lsf} and \ref{subsec:sig}).}
\end{table*}

In this paper, we measure [O\,{\sc ii}]$\lambda$3726/[\oii]$\lambda$3729 ratios to obtain $n_{\rm e}$(\oii).
The separation of the observed wavelengths of the [\oii]$\lambda\lambda$3726,3729 doublet is even shorter than the double of the expected FWHM of the $R\sim2700$ grating, which means that the observed [\oii] doublet is not fully deblended even with the $R\sim2700$ grating.
To deblend the [\oii] doublet, we should determine emission-line profiles of observed galaxies using line-spread functions (LSFs) of NIRSpec and intrinsic velocity dispersions ($\sigma_{\rm int}$) of the observed galaxies.
Section \ref{subsec:lsf} describes how we derive the LSFs, and Section \ref{subsec:sig} explains how we measure $\sigma_{\rm int}$.

\subsection{Line-Spread Function} \label{subsec:lsf}
We obtain the LSFs from a planetary nebula (PN), IRAS-05248-7007.
\tcra{NIRSpec} spectra of the PN were taken during the commissioning process (proposal ID: 1125).
\tcra{The PN was observed with a slit whose width is 0.2 arcsec, which is the same as those of the MSA shutters.}
We use the level-3 spectra available at MAST.
We choose emission lines with high S/N ratios.
We also avoid using emission lines blended with other nearby features.
After the selection, we obtain 4, 11, 5 emission lines for the spectra obtained with the filters F100LP, F170LP, and F290LP, respectively.
The selected emission lines cover wide wavelength ranges of 1.1--1.7, 1.8--2.8, and 3.1--4.5 $\mu$m for the filters F100LP, F170LP, and F290LP, respectively.
\tcra{To obtain the LSF at an arbitrary wavelength, we interpolate the LSF profiles.}

Figure \ref{fig:lsf} summarizes LSFs at different wavelengths for each filter-grating pair.
The obtained LSFs of F100LP-G140H, F170LP-G235H, F290LP-G395H, F100LP-G140M, F170LP-G235M, and F290LP-G395M have median resolution values of $R\simeq3180$, 2800, 3390, 1160, 1120, and 1080 obtained in wavelength ranges of 1.1--1.7, 1.8--2.8, 3.1--4.5, 1.1--1.7, 1.8--2.8, and 3.1--4.5 $\mu$m, respectively.
\tcra{These resolution values are $\sim10$--20\% larger than those of the nominal resolutions of $R\sim2700$ for the high-resolution gratings and $R\sim1000$ for the medium-resolution gratings.}
We also confirm that LSFs of each filter-grating pair become narrower (i.e., better resolution) with longer wavelengths, which is also expected by the NIRSpec instrumental team\footnote{\url{https://jwst-docs.stsci.edu/jwst-near-infrared-spectrograph/nirspec-instrumentation/nirspec-dispersers-and-filters}}.

\tcra{The PN was also observed by VLT/X-shooter with the spectral resolution of $R\sim6500$ (\citealt{Euclid2023}; the spectral properties are available in the Visier atlas\footnote{\url{http://vizier.cfa.harvard.edu/viz-bin/VizieR-3?-source=J/A\%2bA/674/A172/pn-0-11\&-out.max=50\&-out.form=HTML\%20Table\&-out.add=_r\&-out.add=_RAJ,_DEJ\&-sort=_r\&-oc.form=sexa}}).
The observed line widths of the PN are nearly the same as those expected from the spectral resolution of 6500, which suggests that the PN's bulk motions are not resolved with $R\sim6500$.
This also indicates that NIRSpec cannot resolve the PN's bulk motions with the NIRSpec spectral resolution of $\sim1000$ or $\sim2700$.}

\tcra{Nevertheless, the PN may have unresolved bulk motions.
Given that intrinsic velocity dispersions ($\sigma_{\rm int}$) of PNe are typically $\sim 30$ km s$^{-1}$ \citep{Jacob2013}, we may not be able to measure $\sigma_{\rm int}$ values of galaxies with $\sigma_{\rm int}\lesssim30$ km s$^{-1}$.
To put it the other way around, the $\sigma_{\rm int}$ value of 30 km s$^{-1}$ does not change the LSF profiles largely, i.e., the $\sigma_{\rm int}$ accounts for only $\sim1$\% and $\sim8$\% of the observed velocity dispersion of the PN with the $R\sim1000$ and $\sim2700$ gratings, respectively.}

\subsection{Velocity-Dispersion Measurement} \label{subsec:sig}
To obtain $\sigma_{\rm int}$ of the observed galaxies, we fit a Gaussian function convolved with the LSF derived in Section \ref{subsec:lsf}.
To obtain $\sigma_{\rm int}$ with smaller uncertainties, we measure $\sigma_{\rm int}$ of [\oiii]$\lambda$5007, which is the strongest in the observed emission lines.
We conduct the fittings to [\oiii]$\lambda\lambda$4959,5007 simultaneously so that line profiles of the two emission lines are reproduced self-consistently.
We fix the flux ratio [\oiii]$\lambda$5007/[O\,{\sc iii}]$\lambda$4959 to 2.98, which is accurately determined by the Einstein A coefficient \citep{Storey2000}.

We note that our fitting provides $\sigma_{\rm int}\sim0$ km s$^{-1}$ for 2 of the 16 JWST galaxies (\tcra{CEERS\_01173} and \tcra{CEERS\_01539}).
We adopt $\sigma_{\rm int}=0$ km s$^{-1}$ for the 2 JWST galaxies to determine [\oii]-doublet profiles, while true $\sigma_{\rm int}$ values of the 2 JWST galaxies are likely to be larger than 0 km s$^{-1}$ and less than that of the PN used for the LSFs (i.e., $0<\sigma_{\rm int,true}\lesssim30$ km s$^{-1}$; Section \ref{subsec:lsf}).

\subsection{Electron-Density Measurement} \label{subsec:ne}
Using the LSF (Section \ref{subsec:lsf}), we measure fluxes of the [\oii] doublet.
We fit \tcra{3 Gaussian functions convolved by the LSF to [\oii]$\lambda\lambda$3726,3729, and [\oiii]$\lambda5007$ simultaneously with the fixed $z$.}
Figure \ref{fig:o2} shows our fitting results.
We use PyNeb (\citealt{Luridiana2015}; v1.1.15) to obtain $n_{\rm e}$ from the [\oii] doublet.
This method provides $n_{\rm e}$ values consistent with those derived from the updated relation between [\oii]$\lambda3729$/[\oii]$\lambda3726$ and $n_{\rm e}$ reported by \citet{Sanders2016}.
\tcra{In ERO\_06355, GLASS\_150029, GLASS\_160133, and CEERS\_00397, electron temperatures of 11300, 17600, 14800, and 15200 K, respectively, have been measured from the [\oiii]$\lambda4363$ line \citep{Nakajima2023arXiv,Isobe2023c}.
For these 4 galaxies, we use these temperature values when deriving the $n_{\rm e}$ values.
For the other galaxies,} we assume electron temperatures of the JWST galaxies to be 15000 K, which is typical for star-forming dwarf galaxies \citep[e.g.,][]{Izotov2012,Kojima2020,Isobe2022,Nakajima2022b}.
We perform Monte Carlo simulations based on spectral errors to obtain uncertainties of $n_{\rm e}$.
We note that 2 of the 16 JWST galaxies \tcra{(GLASS\_10021 and CEERS\_01449)} show unphysical [\oii]$\lambda3726$/[\oii]$\lambda3729$ ratios, even including the spectral errors into the $n_{\rm e}$ measurements.
Omitting the 2 galaxies, we obtain a final sample consisting of 14 ($=16-2$) galaxies.
Table \ref{tab:ne} lists the derived $n_{\rm e}$ with the uncertainties.

\tcra{Since the $n_{\rm e}$ values based on the $R\sim2700$ observations are slightly lower than those with $R\sim1000$, we have performed a recovery test for the $R\sim1000$ data by making mock spectra around the [\oii]$\lambda\lambda$3726,3729 and [\oiii]$\lambda5007$ lines with a typical $\sigma_{\rm int}$ of 60 km s$^{-1}$ convolved by the LSFs and noises that match typical S/N ratios of these line fluxes.
Conducting line fitting to the mock spectra in the same way as for the scientific data, we have confirmed that our fitting method can reproduce electron density values within at most 3.5\% accuracy even for the $R\sim1000$ spectra.}

We check that different assumptions of electron temperatures from $T_{\rm e}=10000$ to 30000 K can change $n_{\rm e}$ values by at most $\sim20$\% from those calculated by the original assumption of $T_{\rm e}=15000$ K.
This systematics of $n_{\rm e}$ is much smaller than those derived from the spectral errors.

\subsection{$M_{*}$ and SFR} \label{subsec:prop}
\begin{figure}[t]
    \centering
    \includegraphics[width=8.0cm]{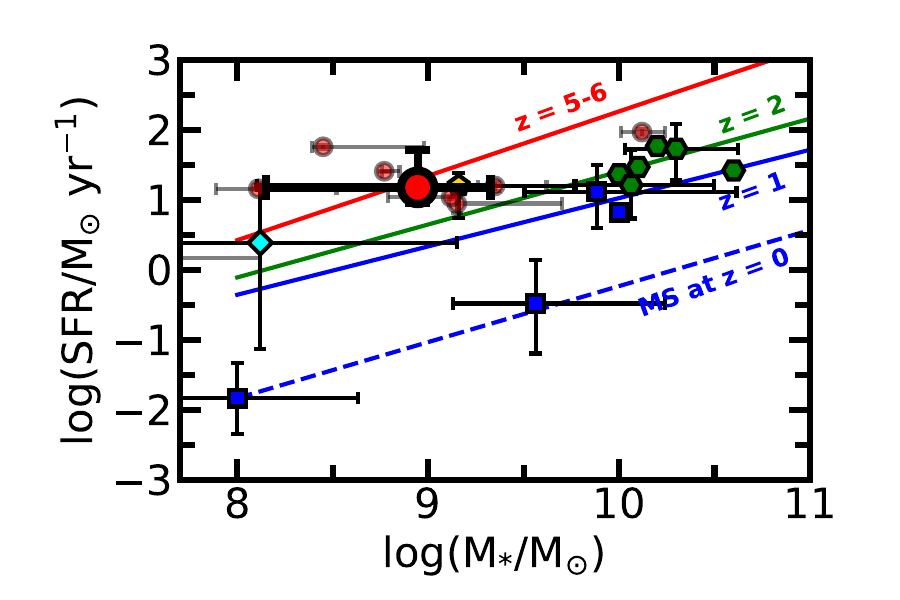}
    \caption{$M_{*}$--SFR relation of the JWST galaxies with the NIRCam photometry at $4<z<9$ represented by the red circles. We also plot median $n_{\rm e}$ values of star-formation main-sequence (SFMS) galaxies with $M_{*}\sim10^{8}$--$10^{10}\ M_{\odot}$ at \tcra{$z\sim0$--1} (blue square; [\oii]- or [\sii]-based; \citealt{Berg2012}; \citealt{Davies2021}; \citealt{Swinbank2019}) and with $M_{*}\sim10^{10}\ M_{\odot}$ at \tcra{$z\sim1$--3} (green hexagon; [\oii]- or [\sii]-based; \citealt{Davies2021}; \tcra{\citealt{Kaasinen2017};} \citealt{Kashino2017}; \citealt{Sanders2016}; \citealt{Steidel2014}). The red solid, green solid, blue solid, and blue dashed lines illustrate SFMSs at $z=5$--6 \citep{Santini2017}, $z=2$ \citep{Speagle2014}, $z=1$ \citep{Speagle2014} and $z=0$ \citep{Chang2015}, respectively. We also show median $n_{\rm e}$ values of high sSFR galaxies with $M_{*}\sim10^{8}\ M_{\odot}$ at $z\sim0$ (cyan diamond; [\sii]-based; \citealt{Berg2022}) and with $M_{*}\sim10^{9}\ M_{\odot}$ at $z\sim2$--3 (yellow pentagon; [\oii]-based; \citealt{Christensen2012a,Christensen2012b,Sanders2016b,Gburek2019}).}
    \label{fig:sfms}
\end{figure}

\begin{figure*}[t]
    \centering
    \includegraphics[width=18.0cm]{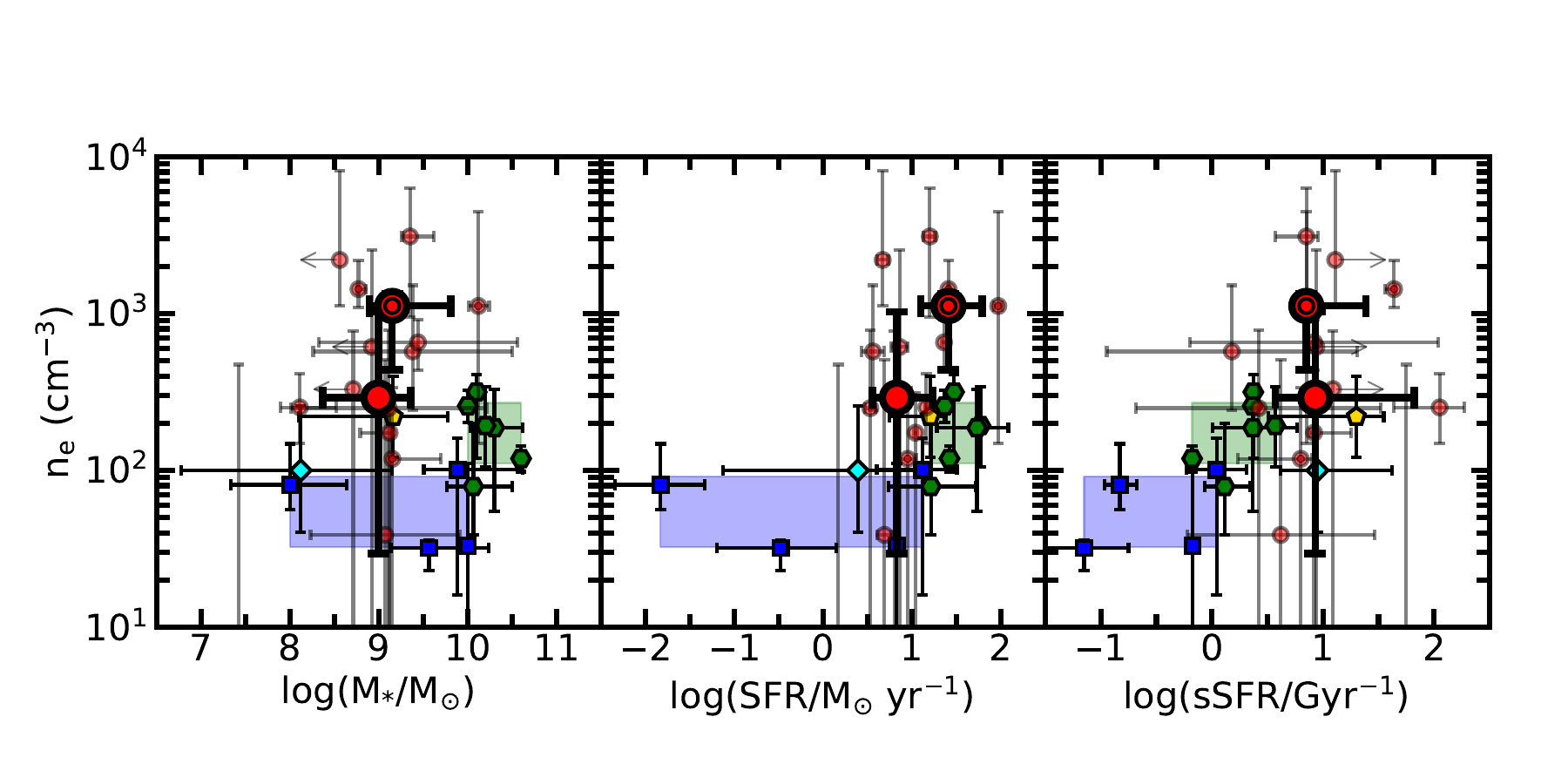}
    \caption{Electron density $n_{\rm e}$ as a function of $M_{*}$ (left), SFR (center), and sSFR (right). The large red circles and the large red double circles denote median and 16th-84th percentiles of the properties of the JWST galaxies at \tcra{$z\sim4$--6} and \tcra{$z\sim7$--9}, respectively. The small red circles and the small red double circle represent the properties of each JWST galaxy at \tcra{$z\sim4$--6} and \tcra{$z\sim7$--9}, respectively. The other symbols are the same as those in Figure \ref{fig:sfms}. The blue and green shaded regions correspond to the 16th-84th percentile ranges of the $n_{\rm e}$ values of the galaxies at \tcra{$z\sim0$--1} \citep{Berg2012,Davies2021,Swinbank2019} and \tcra{$z\sim1$--3} \citep{Davies2021,Kaasinen2017,Kashino2017,Sanders2016,Steidel2014}, respectively. On average, the JWST galaxies have higher $n_{\rm e}$ than those of lower-$z$ galaxies at a given $M_{*}$, SFR, and sSFR.}
    \label{fig:nep}
\end{figure*}

\begin{figure*}[t]
    \centering
    \includegraphics[width=18.0cm]{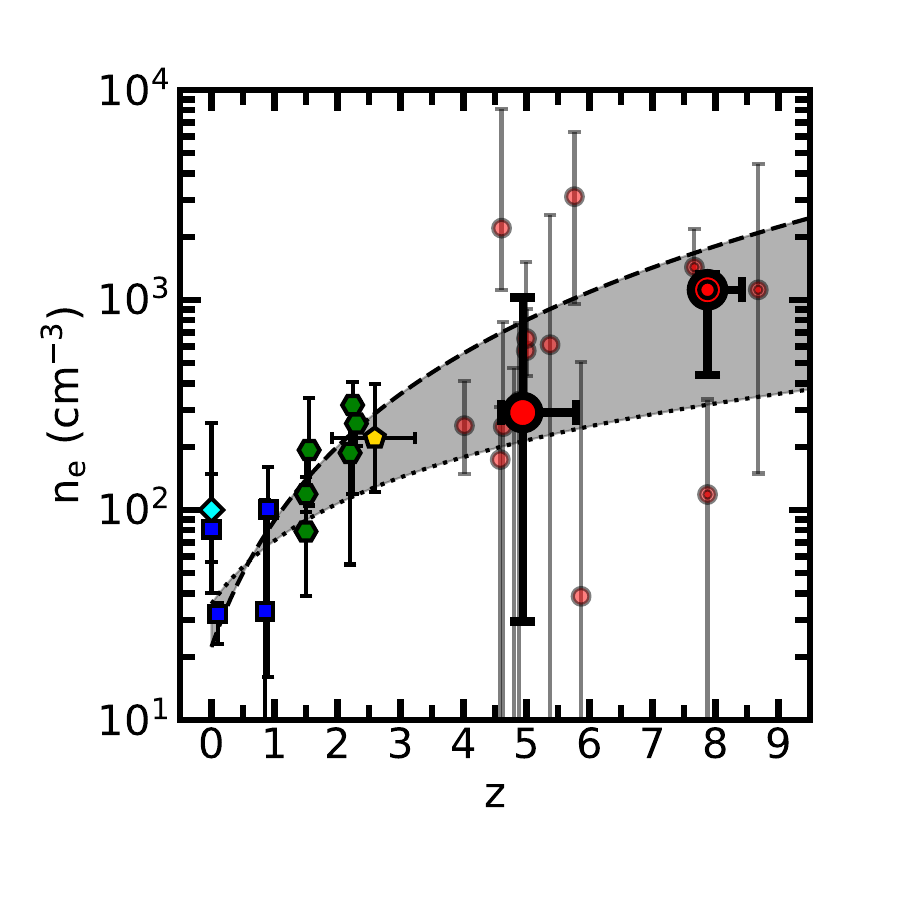}
    \caption{Electron density $n_{\rm e}$ as a function of redshift. All of the measurements in this figure are based on the low-ionization lines (i.e., [\oii] or [\sii]). The symbols are the same as those in Figure \ref{fig:nep}. The dotted and dashed curves represent $n_{\rm e}$ following $\propto(1+z)$ and $\propto(1+z)^{2}$ relations, respectively, which are determined by the galaxies at \tcra{$z\sim0$--3}. The JWST galaxies lie on the gap between the two functions (i.e., $n_{\rm e}\propto(1+z)$ and $\propto(1+z)^{2}$) filled with the gray shaded region.}
    \label{fig:nez}
\end{figure*}

In this paper, we characterize the JWST galaxies with their $M_{*}$ and SFR values.
\tcra{Seven} of the 14 JWST galaxies have JWST/NIRCam photometry at the time of writing.
Regarding these 6 galaxies, we use $M_{*}$ and SFR values derived by \citet{Nakajima2023arXiv}.
\citet{Nakajima2023arXiv} derive $M_{*}$ by conducting SED fitting with prospector \citep{Johnson2021} assuming the \citet{Chabrier2003} IMF, a flexible star formation history, and the \citet{Calzetti2000} dust extinction law, in a similar manner to \citet{Harikane2023}. 
NIRCam photometries of the JWST galaxies are measured by Source Extractor \citep{Bertin1996} as in \citet{Harikane2023}.
The photometries are corrected for gravitational lensing magnifications modeled by \texttt{glafic} \citep{Oguri2010}.
\tcra{We then correct the NIRSpec spectra for slit losses by scaling the spectra to match the NIRCam photometries, where the typical correction factor is about 1.3.}
The SFR values are calculated from slitloss-corrected H$\beta$ luminosities with the NIRCam restframe optical photometry under the assumptions of the \citet{Kennicutt1998} relation and \citet{Chabrier2003} IMF.
Figure \ref{fig:sfms} shows $M_{*}$--SFR relations of the JWST galaxies with the NIRCam photometry.
The median $M_{*}$ and SFR values of the JWST galaxies (large red circle) show that the JWST galaxies generally lie on the star-formation main sequence (SFMS) at $z=5$--6 \citep{Santini2017}.

We estimate $M_{*}$ of the remaining \tcra{7 ($=14-7$)} JWST galaxies from broadband photometry of HST/ACS F606W, F814W, or HST/WFC3 F125W taken from the CANDELS EGS catalog \citep{Stefanon2017}.
We use FLUX\_AUTO values based on Source Extractor \citep{Bertin1996} as total fluxes.
Assuming that the \tcra{7} JWST galaxies lie on the SFMS as well as the \tcra{7} JWST galaxies with the NIRCam photometry, we adopt empirical relations between $M_{*}$ and absolute magnitude of restframe 1500 \AA\ ($M_{\rm UV}$) derived at $z=4$, 5, 6, 7, and 8 (\citealt{Song2016}; based on the \citealt{Salpeter1955} IMF).
We choose the $M_{*}$-$M_{\rm UV}$ relation and the HST photometry both consistent with the redshift of each JWST galaxy.
Note that 3 of the \tcra{7} JWST galaxies do not have $>5\sigma$ detection of HST photometry.
For these 3 JWST galaxies, we calculate upper limits of $M_{*}$ based on $5\sigma$ limiting magnitudes of the HST photometry.
For consistency with $M_{*}$ values derived by \citet{Nakajima2023arXiv}, we rescale the $M_{*}$ values from the \citet{Salpeter1955} to \citet{Chabrier2003} IMF by multiplying 0.61 \citep{Madau2014}.
We check that the $M_{*}$ values based on the HST photometry are consistent with those of \citet{Nakajima2023arXiv} within a $1\sigma$ level.
\tcra{We have propagated errors of the HST photometry and the $M_{*}$-$M_{\rm UV}$ relations to the final stellar mass errors.
The scatters of the $M_{*}$-$M_{\rm UV}$ relations are $\sim0.15$ dex at $z\sim4$ and $\sim0.3$ dex at $z\sim8$, which are smaller than those of our derived stellar mass errors.
This indicates that the stellar mass errors are dominated by the HST photometry.}

We also calculate SFRs of the \tcra{7} JWST galaxies from H$\beta$ luminosities under the assumptions of the \citet{Kennicutt1998} relation and \citet{Chabrier2003} IMF in the same manner as \citet{Nakajima2023arXiv}.
\tcra{Although we do not correct the H$\beta$ luminosities of these \tcra{7} galaxies without the NIRCam photometry for slit losses, the small typical slit-loss correction factor of 1.3 suggests that the slit-loss correction does not change the SFR values much.}
We summarize the $M_{*}$ and SFR values in Table \ref{tab:ne}.

\section{Result and Discussion} \label{sec:res}
Figure \ref{fig:nep} presents $n_{\rm e}$--$M_{*}$ (left), --SFR (center), and --sSFR (right) relations.
The small red circles and the small red double circle in Figure \ref{fig:nep} represent the properties of each JWST galaxy at \tcra{$z\sim4$--6} and \tcra{$z\sim7$--9}, respectively.
The large red circles and the large red double circles denote median and 16th-84th percentiles of the properties of the JWST galaxies at \tcra{$z\sim4$--6} and \tcra{$z\sim7$--9}, respectively.
We find that the median $n_{\rm e}$ values of the JWST galaxies at \tcra{$z\sim4$--6} and \tcra{$z\sim7$--9} are $\sim300$ and $\sim1000$ cm$^{-3}$, respectively.

\tcra{For comparison,} we compile median $n_{\rm e}$ values of SFMS galaxies at \tcra{$z\sim0$--1} (blue square; \citealt{Berg2012,Davies2021,Swinbank2019}) and \tcra{$z\sim1$--3} (green hexagon; [\oii]- or [\sii]-based; \citealt{Davies2021,Kaasinen2017,Kashino2017,Sanders2016,Steidel2014}) whose $M_{*}$--SFR relations are shown in Figure \ref{fig:sfms}.
\tcra{We note that we derive the median $n_{\rm e}$ value of \citet{Berg2012} from the flux ratios of [\sii]$\lambda$6716/[\sii]$\lambda$6731 and $T_{\rm e}$ of 31 galaxies in \citet{Berg2012}.
In addition, the median SFR values of \citet{Davies2021} are read from their figures.}

\tcra{It should be noted that $n_{\rm e}$ values are reported not to be very dependent on $M_{*}$ \citep{Kashino2019} but on SFR and sSFR \citep[e.g.,][]{Shimakawa2015,Jiang2019,Kashino2019,Davies2021} at a given redshift.
To examine how much the $n_{\rm e}$ values change with SFR and sSFR, we plot the 16th-84th percentile ranges of the $n_{\rm e}$ values of the SFMS galaxies at $z\sim0$--1 with the blue shaded region and at $z\sim1$--3 with the green shaded region, respectively, in Figure \ref{fig:nep}.
The blue shaded region indicates that the $n_{\rm e}$ dependencies on $M_{*}$, SFR, and sSFR in the ranges of $\log(M_{*}/M_{\odot})\sim8$--10, $\log(\mathrm{SFR}/M_{\odot}\ \mathrm{yr}^{-1})\sim-2$--1, and $\log(\mathrm{sSFR}/\mathrm{Gyr}^{-1})\sim-1$--0, respectively, are folded in the $n_{\rm e}$ range of $\sim30$--100 cm$^{-3}$.
Note that the sSFR values of the SFMS galaxies at $z\sim0$--1 are lower than those of the JWST galaxies.
The SFMS galaxies at $z\sim1$--3 typically have $n_{\rm e}\sim100$--300, while the parametric ranges examined are relatively narrow, i.e., $\log(M_{*}/M_{\odot})\sim10$--10.5, $\log(\mathrm{SFR}/M_{\odot}\ \mathrm{yr}^{-1})\sim1$--2, and $\log(\mathrm{sSFR}/\mathrm{Gyr}^{-1})\sim0$--0.5.}

\tcra{To cover higher sSFRs that match those of the JWST galaxies,} we also refer to active star-forming galaxies at $z\sim0$ (cyan diamond; \citealt{Berg2022}) and $\sim2$--3 (yellow pentagon; \citealt{Christensen2012a,Christensen2012b,Sanders2016b,Gburek2019})) whose sSFR values are comparable to those of the JWST galaxies (cf. Figure \ref{fig:sfms}).
\tcra{We refer to these galaxies as high sSFR galaxies.
The median values of the high sSFR galaxies at $z\sim0$ and $z\sim2$--3 are $\sim100$ and $\sim200$ cm$^{-3}$, respectively, both of which are comparable to those of the SFMS galaxies at a given redshift.}

\tcra{The left and center panels of Figure \ref{fig:nep} illustrate that the $n_{\rm e}$ values increase from the $z\sim0$--1 to $z\sim1$--3 and $z\sim4$--9 at a given $M_{*}$ and SFR.
This indicates that $n_{\rm e}$ evolves with redshift beyond the range that can be explained by the correlation of $n_{\rm e}$ and SFR.}
\tcra{On the other hand, the sSFR distribution of the JWST galaxies does not overlap with that of the SFMS galaxies at $z\sim0$--3 as shown in the right panel of Figure \ref{fig:nep}, which makes it hard to tell if the $n_{\rm e}$ values evolve with redshift or correlate with sSFR.
However, compared with the high sSFR galaxies at $z\sim0$ (cyan diamond) and $z\sim2$--3 (yellow pentagon), the $n_{\rm e}$ values are likely to increase from the $z\sim0$--1 to $z\sim1$--3 and $z\sim4$--9 at a given sSFR as well.
Note that the $z\sim0$ high sSFR sample consists of 45 galaxies, comparable to those of \citet{Davies2021} in each redshift bin, while the $z\sim2$--3 high sSFR sample consists of only 5 galaxies due to the lack of statistical surveys for $z\sim2$--3 dwarf galaxies.
}

Figure \ref{fig:nez} clearly shows that the $n_{\rm e}$ values typically increase from \tcra{$z\sim0$} to \tcra{$z\sim1$--3} and \tcra{$z\sim4$--9}.
We then investigate the dependence of $n_{\rm e}$ on $z$ by comparing the $n_{\rm e}$ values of the JWST galaxies with an extrapolation of the increasing trend identified in lower-$z$ galaxies.
Assuming that the lower-$z$ galaxies follow $n_{\rm e}\propto(1+z)$ or $\propto(1+z)^{2}$ relation, we fit the 2 functions to the \tcra{$z\sim0$--3} galaxies based on $\chi^{2}$ minimization.
We find that the JWST galaxies at \tcra{$z\sim4$--9} have median $n_{\rm e}$ and $z$ values (large red circles) falling in the gap between the $n_{\rm e}\propto(1+z)$ and $\propto(1+z)^{2}$ relations (gray shaded region), which suggests that galaxies at \tcra{$z\sim0$--9} have an evolutionary relation of $n_{\rm e}\propto(1+z)^{p}$ with $p\sim1$--2.

Although it is not obvious that global galaxy properties impact on the ISM property of $n_{\rm e}$, the $n_{\rm e}\appropto(1+z)^{2}$ relation suggests that nebula densities of high-$z$ star-forming galaxies are generally high due to the compact morphology, i.e., smaller effective radii $r_{\rm e}$ and virial radii $r_{\rm vir}$.
Given the size evolutions of $r_{\rm e}\appropto(1+z)^{-1}$ \citep{Shibuya2015,Ono2023} and $r_{\rm vir}\propto(1+z)^{-1}$ \citep[e.g.,][]{Mo2002}, stellar- and halo-mass densities are expected to be proportional to $r_{\rm e}^{-2}\appropto(1+z)^{2}$ and $r_{\rm vir}^{-2}\propto(1+z)^{2}$ at a given mass, respectively, under the assumption that star-forming galaxies have disks (suggested by their low S{\'e}rsic indices; cf. \citealt{Shibuya2015,Ono2023}) whose heights are constant.
If the stellar-mass and halo-mass densities are proportional to $n_{\rm e}$, $n_{\rm e}$ is approximately proportional to $(1+z)^{2}$.
This suggestion is similar to the conclusion claimed by \citet{Davies2021}.

However, the JWST galaxies generally have $n_{\rm e}$ values slightly lower than those predicted by the $p=2$ evolutionary relation.
\tcra{Although this discrepancy may be explained by the $\sim20$\% uncertainty of the electron temperature (Section \ref{subsec:ne})}, we should think of other factors that can decrease $n_{\rm e}$ values.
One possible explanation is that
\tcra{thermal expansions at higher electron temperatures of high-$z$ metal-poor nebulae result in lower $n_{\rm e}$ values for a given pressure}
\citep{Kewley2019}.
The $p\sim1$--2 evolutionary relation of $n_{\rm e}$ may originate from a combination of the compact morphology and the high electron temperature of high-$z$ galaxies.
\tcra{It should be noted that the JWST galaxies have $M_{*}$ values $\sim1$ dex smaller than those of the SFMS galaxies at $z\sim0$--3.
Since the observed size-mass relation is a positive trend, $z\sim0$--3 galaxies with similar $M_{*}$ to the JWST galaxies may have larger $n_{\rm e}$ values than the current SFMS samples.
This could lead to an even flatter redshift evolution of $n_{\rm e}$.}

\tcra{It should also be noted} that there is a possibility that the relatively-high S/N criterion for the [\oii] doublet (Section \ref{sec:datsam}) causes a bias towards lower [\oiii]$\lambda$5007/[\oii]$\lambda\lambda$3727,3729 ratios (O32, hereafter).
\tcra{This potential bias leads to ionization parameters $q$ lower \citep[e.g.,][]{Kewley2002}, which leads to high $n_{\rm e}$ evidenced with $q\propto n_{\rm H}^{-1}$, where $n_{\rm H}$ is a hydrogen number density roughly proportional to $n_{\rm e}$.
However, the JWST galaxies already have a high median O32 value of $\simeq5$, which indicates that the JWST galaxies are not biased towards low O32.}

\section{Summary} \label{sec:sum}
We present electron densities of singly-ionized regions $n_{\rm e}$(\oii) in the inter-stellar medium (ISM) of star-formation main-sequence galaxies at $z=4$--9 observed by JWST/NIRSpec ERO, GLASS, and CEERS programs.
Deriving line-spread functions of the NIRSpec instrument from in-flight calibration data, we securely measure [\oii]$\lambda$3726/[\oii]$\lambda$3729 ratios of 14 galaxies at $z=4.02$--8.68.
We find that the 14 galaxies have $n_{\rm e}\gtrsim300$ cm$^{-3}$ significantly higher than those of low-$z$ galaxies at a given stellar mass, star-formation rate (SFR), and specific SFR (Figure \ref{fig:nep}).
We also identify an increase in typical $n_{\rm e}$ values from $z=0$ to $z=1$--3 and $z=4$--9, which is approximated by $n_{\rm e}\propto(1+z)^{p}$ with $p\sim1$--2 (Figure \ref{fig:nez}).
Although it is not obvious that the ISM property of $n_{\rm e}$ is influenced by global galaxy properties, the $p\sim1$--2 evolutionary relation can be explained by a combination of the compact morphology and the reduction of $n_{\rm e}$ due to high electron temperatures of high-$z$ metal-poor nebulae.
\\

We thank M. Oguri for providing his lens models.
We also thank H. Yajima and H. Fukushima for having useful discussions.
We are grateful to staff of the James Webb Space Telescope Help Desk for letting us know useful information.
This work is based on observations made with the NASA/ESA/CSA James Webb Space Telescope.
\tcra{Some of the data presented in this paper were obtained from the Mikulski Archive for Space Telescopes (MAST) at the Space Telescope Science Institute, which is operated by the Association of Universities for Research in Astronomy, Inc., under NASA contract NAS 5-03127 for JWST. The specific observations analyzed can be accessed via \dataset[10.17909/qfxm-y747]{https://doi.org/10.17909/qfxm-y747}.}
These observations are associated with programs 1125, 2736, 1324, and 1345.
The authors acknowledge the ERO, GLASS, and CEERS teams led by Klaus M. Pontoppidan, Tommaso Treu, and Steven L. Finkelstein, respectively, for developing their observing programs with a zero-exclusive-access period.
This work is based on observations taken by the CANDELS Multi-Cycle Treasury Program with the NASA/ESA HST, which is operated by the Association of Universities for Research in Astronomy, Inc., under NASA contract NAS5-26555.
This work was supported by the joint research program of the Institute for Cosmic Ray Research (ICRR), University of Tokyo. 
Y.I., K.N., and Y.H. are supported by JSPS KAKENHI Grant Nos. 21J20785, 20K22373, and 21K13953, respectively.
This paper is supported by World Premier International Research Center Initiative (WPI Initiative), MEXT, Japan, as well as the joint research program of the Institute of Cosmic Ray Research (ICRR), the University of Tokyo. This work is supported by KAKENHI (19H00697, 20H00180, and 21H04467) Grant-in-Aid for Scientific Research (A) through the Japan Society for the Promotion of Science. 
This research was supported by a grant from the Hayakawa Satio Fund awarded by the Astronomical Society of Japan.
\software{astropy \citep{Astropy2013,Astropy2018}, PyNeb \citep{Luridiana2015}, prospector \citep{Johnson2021}, glafic \citep{Oguri2010}, Source Extractor \citep{Bertin1996}}

%

\bibliography{library}

\begin{thebibliography}{}
\expandafter\ifx\csname natexlab\endcsname\relax\def\natexlab#1{#1}\fi
\providecommand{\url}[1]{\href{#1}{#1}}
\providecommand{\dodoi}[1]{doi:~\href{http://doi.org/#1}{\nolinkurl{#1}}}
\providecommand{\doeprint}[1]{\href{http://ascl.net/#1}{\nolinkurl{http://ascl.net/#1}}}
\providecommand{\doarXiv}[1]{\href{https://arxiv.org/abs/#1}{\nolinkurl{https://arxiv.org/abs/#1}}}

\bibitem[{{Astropy Collaboration} {et~al.}(2013){Astropy Collaboration},
  {Robitaille}, {Tollerud}, {Greenfield}, {Droettboom}, {Bray}, {Aldcroft},
  {Davis}, {Ginsburg}, {Price-Whelan}, {Kerzendorf}, {Conley}, {Crighton},
  {Barbary}, {Muna}, {Ferguson}, {Grollier}, {Parikh}, {Nair}, {Unther},
  {Deil}, {Woillez}, {Conseil}, {Kramer}, {Turner}, {Singer}, {Fox}, {Weaver},
  {Zabalza}, {Edwards}, {Azalee Bostroem}, {Burke}, {Casey}, {Crawford},
  {Dencheva}, {Ely}, {Jenness}, {Labrie}, {Lim}, {Pierfederici}, {Pontzen},
  {Ptak}, {Refsdal}, {Servillat}, \& {Streicher}}]{Astropy2013}
{Astropy Collaboration}, {Robitaille}, T.~P., {Tollerud}, E.~J., {et~al.} 2013,
  \aap, 558, A33, \dodoi{10.1051/0004-6361/201322068}

\bibitem[{{Berg} {et~al.}(2021){Berg}, {Chisholm}, {Erb}, {Skillman}, {Pogge},
  \& {Olivier}}]{Berg2021}
{Berg}, D.~A., {Chisholm}, J., {Erb}, D.~K., {et~al.} 2021, \apj, 922, 170,
  \dodoi{10.3847/1538-4357/ac141b}

\bibitem[{{Berg} {et~al.}(2012){Berg}, {Skillman}, {Marble}, {van Zee},
  {Engelbracht}, {Lee}, {Kennicutt}, {Calzetti}, {Dale}, \&
  {Johnson}}]{Berg2012}
{Berg}, D.~A., {Skillman}, E.~D., {Marble}, A.~R., {et~al.} 2012, \apj, 754,
  98, \dodoi{10.1088/0004-637X/754/2/98}

\bibitem[{{Berg} {et~al.}(2022){Berg}, {James}, {King}, {McDonald}, {Chen},
  {Chisholm}, {Heckman}, {Martin}, {Stark}, {Aloisi}, {Amor{\'\i}n},
  {Arellano-C{\'o}rdova}, {Bayliss}, {Bordoloi}, {Brinchmann}, {Charlot},
  {Chevallard}, {Clark}, {Erb}, {Feltre}, {Gronke}, {Hayes}, {Henry},
  {Hernandez}, {Jaskot}, {Jones}, {Kewley}, {Kumari}, {Leitherer}, {Llerena},
  {Maseda}, {Mingozzi}, {Nanayakkara}, {Ouchi}, {Plat}, {Pogge},
  {Ravindranath}, {Rigby}, {Sanders}, {Scarlata}, {Senchyna}, {Skillman},
  {Steidel}, {Strom}, {Sugahara}, {Wilkins}, {Wofford}, {Xu}, \& {Classy
  Team}}]{Berg2022}
{Berg}, D.~A., {James}, B.~L., {King}, T., {et~al.} 2022, \apjs, 261, 31,
  \dodoi{10.3847/1538-4365/ac6c03}

\bibitem[{{Bertin} \& {Arnouts}(1996)}]{Bertin1996}
{Bertin}, E., \& {Arnouts}, S. 1996, \aaps, 117, 393,
  \dodoi{10.1051/aas:1996164}

\bibitem[{Calzetti {et~al.}(2000)Calzetti, Armus, Bohlin, Kinney, Koornneef, \&
  Storchi‐Bergmann}]{Calzetti2000}
Calzetti, D., Armus, L., Bohlin, R.~C., {et~al.} 2000, ApJ, 533, 682,
  \dodoi{10.1086/308692}

\bibitem[{{Chabrier}(2003)}]{Chabrier2003}
{Chabrier}, G. 2003, \pasp, 115, 763, \dodoi{10.1086/376392}

\bibitem[{{Chang} {et~al.}(2015){Chang}, {van der Wel}, {da Cunha}, \&
  {Rix}}]{Chang2015}
{Chang}, Y.-Y., {van der Wel}, A., {da Cunha}, E., \& {Rix}, H.-W. 2015, \apjs,
  219, 8, \dodoi{10.1088/0067-0049/219/1/8}

\bibitem[{{Christensen} {et~al.}(2012{\natexlab{a}}){Christensen}, {Richard},
  {Hjorth}, {Milvang-Jensen}, {Laursen}, {Limousin}, {Dessauges-Zavadsky},
  {Grillo}, \& {Ebeling}}]{Christensen2012a}
{Christensen}, L., {Richard}, J., {Hjorth}, J., {et~al.} 2012{\natexlab{a}},
  \mnras, 427, 1953, \dodoi{10.1111/j.1365-2966.2012.22006.x}

\bibitem[{{Christensen} {et~al.}(2012{\natexlab{b}}){Christensen}, {Laursen},
  {Richard}, {Hjorth}, {Milvang-Jensen}, {Dessauges-Zavadsky}, {Limousin},
  {Grillo}, \& {Ebeling}}]{Christensen2012b}
{Christensen}, L., {Laursen}, P., {Richard}, J., {et~al.} 2012{\natexlab{b}},
  \mnras, 427, 1973, \dodoi{10.1111/j.1365-2966.2012.22007.x}

\bibitem[{{Davies} {et~al.}(2021){Davies}, {F{\"o}rster Schreiber}, {Genzel},
  {Shimizu}, {Davies}, {Schruba}, {Tacconi}, {{\"U}bler}, {Wisnioski}, {Wuyts},
  {Fossati}, {Herrera-Camus}, {Lutz}, {Mendel}, {Naab}, {Price}, {Renzini},
  {Wilman}, {Beifiori}, {Belli}, {Burkert}, {Chan}, {Contursi}, {Fabricius},
  {Lee}, {Saglia}, \& {Sternberg}}]{Davies2021}
{Davies}, R.~L., {F{\"o}rster Schreiber}, N.~M., {Genzel}, R., {et~al.} 2021,
  \apj, 909, 78, \dodoi{10.3847/1538-4357/abd551}

\bibitem[{{Euclid Collaboration} {et~al.}(2023){Euclid Collaboration},
  {Paterson}, {Schirmer}, {Copin}, {Cuillandre}, {Gillard}, {Guti{\'e}rrez
  Soto}, {Guzzo}, {Hoekstra}, {Kitching}, {Paltani}, {Percival}, {Scodeggio},
  {Stanghellini}, {Appleton}, {Laureijs}, {Mellier}, {Aghanim}, {Altieri},
  {Amara}, {Auricchio}, {Baldi}, {Bender}, {Bodendorf}, {Bonino}, {Branchini},
  {Brescia}, {Brinchmann}, {Camera}, {Capobianco}, {Carbone}, {Carretero},
  {Castander}, {Castellano}, {Cavuoti}, {Cimatti}, {Cledassou}, {Congedo},
  {Conselice}, {Conversi}, {Corcione}, {Courbin}, {Da Silva}, {Degaudenzi},
  {Dinis}, {Douspis}, {Dubath}, {Dupac}, {Ferriol}, {Frailis}, {Franceschi},
  {Fumana}, {Galeotta}, {Garilli}, {Gillis}, {Giocoli}, {Grazian}, {Grupp},
  {Haugan}, {Holmes}, {Hornstrup}, {Hudelot}, {Jahnke}, {K{\"u}mmel},
  {Kiessling}, {Kilbinger}, {Kohley}, {Kubik}, {Kunz}, {Kurki-Suonio},
  {Ligori}, {Lilje}, {Lloro}, {Maiorano}, {Mansutti}, {Marggraf}, {Markovic},
  {Marulli}, {Massey}, {Medinaceli}, {Mei}, {Meneghetti}, {Meylan}, {Moresco},
  {Moscardini}, {Nakajima}, {Niemi}, {Nightingale}, {Nutma}, {Padilla},
  {Pasian}, {Pedersen}, {Polenta}, {Poncet}, {Popa}, {Raison}, {Renzi},
  {Rhodes}, {Riccio}, {Rix}, {Romelli}, {Roncarelli}, {Rossetti}, {Saglia},
  {Sartoris}, {Schneider}, {Secroun}, {Seidel}, {Serrano}, {Sirignano},
  {Sirri}, {Skottfelt}, {Stanco}, {Tallada-Cresp{\'\i}}, {Taylor}, {Tereno},
  {Toledo-Moreo}, {Torradeflot}, {Tutusaus}, {Valenziano}, {Vassallo}, {Wang},
  {Weller}, {Zamorani}, {Zoubian}, {Andreon}, {Bardelli}, {Bozzo},
  {Colodro-Conde}, {Di Ferdinando}, {Farina}, {Graci{\'a}-Carpio},
  {Keih{\"a}nen}, {Lindholm}, {Maino}, {Mauri}, {Scottez}, {Tenti}, {Zucca},
  {Akrami}, {Baccigalupi}, {Ballardini}, {Biviano}, {Borlaff}, {Burigana},
  {Cabanac}, {Cappi}, {Carvalho}, {Casas}, {Castignani}, {Castro}, {Chambers},
  {Cooray}, {Coupon}, {Courtois}, {Davini}, {De Lucia}, {Desprez}, {Escartin},
  {Escoffier}, {Ferrero}, {Gabarra}, {Garcia-Bellido}, {George}, {Giacomini},
  {Gozaliasl}, {Hildebrandt}, {Hook}, {Kajava}, {Kansal}, {Kirkpatrick},
  {Legrand}, {Loureiro}, {Magliocchetti}, {Mainetti}, {Maoli}, {Marcin},
  {Martinelli}, {Martinet}, {Martins}, {Matthew}, {Maurin}, {Metcalf},
  {Monaco}, {Morgante}, {Nadathur}, {Patrizii}, {Pollack}, {Porciani},
  {Potter}, {P{\"o}ntinen}, {S{\'a}nchez}, {Sakr}, {Schneider}, {Sefusatti},
  {Sereno}, {Shulevski}, {Stadel}, {Steinwagner}, {Valieri}, {Valiviita},
  {Veropalumbo}, {Viel}, \& {Zinchenko}}]{Euclid2023}
{Euclid Collaboration}, {Paterson}, K., {Schirmer}, M., {et~al.} 2023, \aap,
  674, A172, \dodoi{10.1051/0004-6361/202346252}

\bibitem[{Feltre {et~al.}(2016)Feltre, Charlot, \& Gutkin}]{Feltre2016}
Feltre, A., Charlot, S., \& Gutkin, J. 2016, MNRAS, 456, 3354,
  \dodoi{10.1093/mnras/stv2794}

\bibitem[{{Finkelstein} {et~al.}(2023){Finkelstein}, {Bagley}, {Ferguson},
  {Wilkins}, {Kartaltepe}, {Papovich}, {Yung}, {Haro}, {Behroozi}, {Dickinson},
  {Kocevski}, {Koekemoer}, {Larson}, {Le Bail}, {Morales},
  {P{\'e}rez-Gonz{\'a}lez}, {Burgarella}, {Dav{\'e}}, {Hirschmann},
  {Somerville}, {Wuyts}, {Bromm}, {Casey}, {Fontana}, {Fujimoto}, {Gardner},
  {Giavalisco}, {Grazian}, {Grogin}, {Hathi}, {Hutchison}, {Jha}, {Jogee},
  {Kewley}, {Kirkpatrick}, {Long}, {Lotz}, {Pentericci}, {Pierel}, {Pirzkal},
  {Ravindranath}, {Ryan}, {Trump}, {Yang}, {Bhatawdekar}, {Bisigello}, {Buat},
  {Calabr{\`o}}, {Castellano}, {Cleri}, {Cooper}, {Croton}, {Daddi}, {Dekel},
  {Elbaz}, {Franco}, {Gawiser}, {Holwerda}, {Huertas-Company}, {Jaskot},
  {Leung}, {Lucas}, {Mobasher}, {Pandya}, {Tacchella}, {Weiner}, \&
  {Zavala}}]{Finkelstein2023}
{Finkelstein}, S.~L., {Bagley}, M.~B., {Ferguson}, H.~C., {et~al.} 2023, \apjl,
  946, L13, \dodoi{10.3847/2041-8213/acade4}

\bibitem[{{Fujimoto} {et~al.}(2022){Fujimoto}, {Ouchi}, {Nakajima}, {Harikane},
  {Isobe}, {Brammer}, {Oguri}, {Gim{\'e}nez-Arteaga}, {Heintz}, {Kokorev},
  {Bauer}, {Ferrara}, {Kojima}, {Lagos}, {Laura}, {Schaerer}, {Shimasaku},
  {Hatsukade}, {Kohno}, {Sun}, {Valentino}, {Watson}, {Fudamoto}, {Inoue},
  {Gonz{\'a}lez-L{\'o}pez}, {Koekemoer}, {Knudsen}, {Lee}, {Magdis}, {Richard},
  {Strait}, {Sugahara}, {Tamura}, {Toft}, {Umehata}, \&
  {Walth}}]{Fujimoto2022arXiv}
{Fujimoto}, S., {Ouchi}, M., {Nakajima}, K., {et~al.} 2022, arXiv e-prints,
  arXiv:2212.06863.
\newblock \doarXiv{2212.06863}

\bibitem[{{Gburek} {et~al.}(2019){Gburek}, {Siana}, {Alavi}, {Emami},
  {Richard}, {Freeman}, {Stark}, {Snapp-Kolas}, \& {Lucero}}]{Gburek2019}
{Gburek}, T., {Siana}, B., {Alavi}, A., {et~al.} 2019, \apj, 887, 168,
  \dodoi{10.3847/1538-4357/ab5713}

\bibitem[{{Harikane} {et~al.}(2020){Harikane}, {Ouchi}, {Inoue}, {Matsuoka},
  {Tamura}, {Bakx}, {Fujimoto}, {Moriwaki}, {Ono}, {Nagao}, {Tadaki}, {Kojima},
  {Shibuya}, {Egami}, {Ferrara}, {Gallerani}, {Hashimoto}, {Kohno}, {Matsuda},
  {Matsuo}, {Pallottini}, {Sugahara}, \& {Vallini}}]{Harikane2020}
{Harikane}, Y., {Ouchi}, M., {Inoue}, A.~K., {et~al.} 2020, \apj, 896, 93,
  \dodoi{10.3847/1538-4357/ab94bd}

\bibitem[{{Harikane} {et~al.}(2023){Harikane}, {Ouchi}, {Oguri}, {Ono},
  {Nakajima}, {Isobe}, {Umeda}, {Mawatari}, \& {Zhang}}]{Harikane2023}
{Harikane}, Y., {Ouchi}, M., {Oguri}, M., {et~al.} 2023, \apjs, 265, 5,
  \dodoi{10.3847/1538-4365/acaaa9}

\bibitem[{{Isobe} {et~al.}(2022){Isobe}, {Ouchi}, {Suzuki}, {Moriya},
  {Nakajima}, {Nomoto}, {Rauch}, {Harikane}, {Kojima}, {Ono}, {Fujimoto},
  {Inoue}, {Kim}, {Komiyama}, {Kusakabe}, {Lee}, {Maseda}, {Matthee},
  {Michel-Dansac}, {Nagao}, {Nanayakkara}, {Nishigaki}, {Onodera}, {Sugahara},
  \& {Xu}}]{Isobe2022}
{Isobe}, Y., {Ouchi}, M., {Suzuki}, A., {et~al.} 2022, \apj, 925, 111,
  \dodoi{10.3847/1538-4357/ac3509}

\bibitem[{{Isobe} {et~al.}(2023){Isobe}, {Ouchi}, {Tominaga}, {Watanabe},
  {Nakajima}, {Umeda}, {Yajima}, {Harikane}, {Fukushima}, {Xu}, {Ono}, \&
  {Zhang}}]{Isobe2023c}
{Isobe}, Y., {Ouchi}, M., {Tominaga}, N., {et~al.} 2023, arXiv e-prints,
  arXiv:2307.00710, \dodoi{10.48550/arXiv.2307.00710}

\bibitem[{{Izotov} {et~al.}(2012){Izotov}, {Thuan}, \& {Guseva}}]{Izotov2012}
{Izotov}, Y.~I., {Thuan}, T.~X., \& {Guseva}, N.~G. 2012, \aap, 546, A122,
  \dodoi{10.1051/0004-6361/201219733}

\bibitem[{{Jacob} {et~al.}(2013){Jacob}, {Sch{\"o}nberner}, \&
  {Steffen}}]{Jacob2013}
{Jacob}, R., {Sch{\"o}nberner}, D., \& {Steffen}, M. 2013, \aap, 558, A78,
  \dodoi{10.1051/0004-6361/201321532}

\bibitem[{Jiang {et~al.}(2019)Jiang, Dekel, Freundlich, Romanowsky, Dutton,
  Macci{\`{o}}, \& {Di Cintio}}]{Jiang2019}
Jiang, F., Dekel, A., Freundlich, J., {et~al.} 2019, MNRAS, 487, 5272,
  \dodoi{10.1093/mnras/stz1499}

\bibitem[{{Jiang} {et~al.}(2021){Jiang}, {Kashikawa}, {Wang}, {Walth}, {Ho},
  {Cai}, {Egami}, {Fan}, {Ito}, {Liang}, {Schaerer}, \& {Stark}}]{Jiang2021}
{Jiang}, L., {Kashikawa}, N., {Wang}, S., {et~al.} 2021, Nature Astronomy, 5,
  256, \dodoi{10.1038/s41550-020-01275-y}

\bibitem[{{Johnson} {et~al.}(2021){Johnson}, {Leja}, {Conroy}, \&
  {Speagle}}]{Johnson2021}
{Johnson}, B.~D., {Leja}, J., {Conroy}, C., \& {Speagle}, J.~S. 2021, \apjs,
  254, 22, \dodoi{10.3847/1538-4365/abef67}

\bibitem[{{Kaasinen} {et~al.}(2017){Kaasinen}, {Bian}, {Groves}, {Kewley}, \&
  {Gupta}}]{Kaasinen2017}
{Kaasinen}, M., {Bian}, F., {Groves}, B., {Kewley}, L.~J., \& {Gupta}, A. 2017,
  \mnras, 465, 3220, \dodoi{10.1093/mnras/stw2827}

\bibitem[{{Kashino} \& {Inoue}(2019)}]{Kashino2019}
{Kashino}, D., \& {Inoue}, A.~K. 2019, \mnras, 486, 1053,
  \dodoi{10.1093/mnras/stz881}

\bibitem[{{Kashino} {et~al.}(2017){Kashino}, {Silverman}, {Sanders},
  {Kartaltepe}, {Daddi}, {Renzini}, {Valentino}, {Rodighiero}, {Juneau},
  {Kewley}, {Zahid}, {Arimoto}, {Nagao}, {Chu}, {Sugiyama}, {Civano}, {Ilbert},
  {Kajisawa}, {Le F{\`e}vre}, {Maier}, {Masters}, {Miyaji}, {Onodera},
  {Puglisi}, \& {Taniguchi}}]{Kashino2017}
{Kashino}, D., {Silverman}, J.~D., {Sanders}, D., {et~al.} 2017, \apj, 835, 88,
  \dodoi{10.3847/1538-4357/835/1/88}

\bibitem[{Kennicutt(1998)}]{Kennicutt1998}
Kennicutt, R.~C. 1998, ARA{\&}A, 36, 189,
  \dodoi{10.1146/annurev.astro.36.1.189}

\bibitem[{{Kewley} \& {Dopita}(2002)}]{Kewley2002}
{Kewley}, L.~J., \& {Dopita}, M.~A. 2002, \apjs, 142, 35,
  \dodoi{10.1086/341326}

\bibitem[{{Kewley} {et~al.}(2019){Kewley}, {Nicholls}, {Sutherland}, {Rigby},
  {Acharya}, {Dopita}, \& {Bayliss}}]{Kewley2019}
{Kewley}, L.~J., {Nicholls}, D.~C., {Sutherland}, R., {et~al.} 2019, \apj, 880,
  16, \dodoi{10.3847/1538-4357/ab16ed}

\bibitem[{{Killi} {et~al.}(2023){Killi}, {Watson}, {Fujimoto}, {Akins},
  {Knudsen}, {Richard}, {Harikane}, {Rigopoulou}, {Rizzo}, {Ginolfi},
  {Popping}, \& {Kokorev}}]{Killi2023}
{Killi}, M., {Watson}, D., {Fujimoto}, S., {et~al.} 2023, \mnras, 521, 2526,
  \dodoi{10.1093/mnras/stad687}

\bibitem[{Kojima {et~al.}(2020)Kojima, Ouchi, Rauch, Ono, Nakajima, Isobe,
  Fujimoto, Harikane, Hashimoto, Hayashi, Komiyama, Kusakabe, Kim, Lee, Mukae,
  Nagao, Onodera, Shibuya, Sugahara, Umemura, \& Yabe}]{Kojima2020}
Kojima, T., Ouchi, M., Rauch, M., {et~al.} 2020, ApJ, 898, 142,
  \dodoi{10.3847/1538-4357/aba047}

\bibitem[{{Luridiana} {et~al.}(2015){Luridiana}, {Morisset}, \&
  {Shaw}}]{Luridiana2015}
{Luridiana}, V., {Morisset}, C., \& {Shaw}, R.~A. 2015, \aap, 573, A42,
  \dodoi{10.1051/0004-6361/201323152}

\bibitem[{Madau \& Dickinson(2014)}]{Madau2014}
Madau, P., \& Dickinson, M. 2014, ARA{\&}A, 52, 415,
  \dodoi{10.1146/annurev-astro-081811-125615}

\bibitem[{{Matsumoto} {et~al.}(2022){Matsumoto}, {Ouchi}, {Nakajima},
  {Kawasaki}, {Murai}, {Motohara}, {Harikane}, {Ono}, {Kushibiki}, {Koyama},
  {Aoyama}, {Konishi}, {Takahashi}, {Isobe}, {Umeda}, {Sugahara}, {Onodera},
  {Nagamine}, {Kusakabe}, {Hirai}, {Moriya}, {Shibuya}, {Komiyama},
  {Fukushima}, {Fujimoto}, {Hattori}, {Hayashi}, {Inoue}, {Kikuchihara},
  {Kojima}, {Koyama}, {Lee}, {Mawatari}, {Miyata}, {Nagao}, {Ozaki}, {Rauch},
  {Saito}, {Suzuki}, {Takeuchi}, {Umemura}, {Xu}, {Yabe}, {Zhang}, \&
  {Yoshii}}]{Matsumoto2022}
{Matsumoto}, A., {Ouchi}, M., {Nakajima}, K., {et~al.} 2022, \apj, 941, 167,
  \dodoi{10.3847/1538-4357/ac9ea1}

\bibitem[{{Mo} \& {White}(2002)}]{Mo2002}
{Mo}, H.~J., \& {White}, S.~D.~M. 2002, \mnras, 336, 112,
  \dodoi{10.1046/j.1365-8711.2002.05723.x}

\bibitem[{{Nagao} {et~al.}(2012){Nagao}, {Maiolino}, {De Breuck}, {Caselli},
  {Hatsukade}, \& {Saigo}}]{Nagao2012}
{Nagao}, T., {Maiolino}, R., {De Breuck}, C., {et~al.} 2012, \aap, 542, L34,
  \dodoi{10.1051/0004-6361/201219518}

\bibitem[{Nakajima \& Ouchi(2014)}]{Nakajima2014}
Nakajima, K., \& Ouchi, M. 2014, MNRAS, 442, 900, \dodoi{10.1093/mnras/stu902}

\bibitem[{{Nakajima} {et~al.}(2023){Nakajima}, {Ouchi}, {Isobe}, {Harikane},
  {Zhang}, {Ono}, {Umeda}, \& {Oguri}}]{Nakajima2023arXiv}
{Nakajima}, K., {Ouchi}, M., {Isobe}, Y., {et~al.} 2023, arXiv e-prints,
  arXiv:2301.12825, \dodoi{10.48550/arXiv.2301.12825}

\bibitem[{{Nakajima} {et~al.}(2022){Nakajima}, {Ouchi}, {Xu}, {Rauch},
  {Harikane}, {Nishigaki}, {Isobe}, {Kusakabe}, {Nagao}, {Ono}, {Onodera},
  {Sugahara}, {Kim}, {Komiyama}, {Lee}, \& {Zahedy}}]{Nakajima2022b}
{Nakajima}, K., {Ouchi}, M., {Xu}, Y., {et~al.} 2022, \apjs, 262, 3,
  \dodoi{10.3847/1538-4365/ac7710}

\bibitem[{{Oguri}(2010)}]{Oguri2010}
{Oguri}, M. 2010, \pasj, 62, 1017, \dodoi{10.1093/pasj/62.4.1017}

\bibitem[{{Ono} {et~al.}(2023){Ono}, {Harikane}, {Ouchi}, {Yajima}, {Abe},
  {Isobe}, {Shibuya}, {Wise}, {Zhang}, {Nakajima}, \& {Umeda}}]{Ono2023}
{Ono}, Y., {Harikane}, Y., {Ouchi}, M., {et~al.} 2023, \apj, 951, 72,
  \dodoi{10.3847/1538-4357/acd44a}

\bibitem[{{Pontoppidan} {et~al.}(2022){Pontoppidan}, {Barrientes}, {Blome},
  {Braun}, {Brown}, {Carruthers}, {Coe}, {DePasquale}, {Espinoza}, {Marin},
  {Gordon}, {Henry}, {Hustak}, {James}, {Jenkins}, {Koekemoer}, {LaMassa},
  {Law}, {Lockwood}, {Moro-Martin}, {Mullally}, {Pagan}, {Player}, {Proffitt},
  {Pulliam}, {Ramsay}, {Ravindranath}, {Reid}, {Robberto}, {Sabbi}, {Ubeda},
  {Balogh}, {Flanagan}, {Gardner}, {Hasan}, {Meinke}, \&
  {Nota}}]{Pontoppidan2022}
{Pontoppidan}, K.~M., {Barrientes}, J., {Blome}, C., {et~al.} 2022, \apjl, 936,
  L14, \dodoi{10.3847/2041-8213/ac8a4e}

\bibitem[{{Salpeter}(1955)}]{Salpeter1955}
{Salpeter}, E.~E. 1955, \apj, 121, 161, \dodoi{10.1086/145971}

\bibitem[{{Sanders} {et~al.}(2016{\natexlab{a}}){Sanders}, {Shapley}, {Kriek},
  {Reddy}, {Freeman}, {Coil}, {Siana}, {Mobasher}, {Shivaei}, {Price}, \& {de
  Groot}}]{Sanders2016}
{Sanders}, R.~L., {Shapley}, A.~E., {Kriek}, M., {et~al.} 2016{\natexlab{a}},
  \apj, 816, 23, \dodoi{10.3847/0004-637X/816/1/23}

\bibitem[{{Sanders} {et~al.}(2016{\natexlab{b}}){Sanders}, {Shapley}, {Kriek},
  {Reddy}, {Freeman}, {Coil}, {Siana}, {Mobasher}, {Shivaei}, {Price}, \& {de
  Groot}}]{Sanders2016b}
---. 2016{\natexlab{b}}, \apjl, 825, L23, \dodoi{10.3847/2041-8205/825/2/L23}

\bibitem[{Santini {et~al.}(2017)Santini, Fontana, Castellano, Criscienzo,
  Merlin, Amorin, Cullen, Daddi, Dickinson, Dunlop, Grazian, Lamastra, McLure,
  Micha{\l}owski, Pentericci, \& Shu}]{Santini2017}
Santini, P., Fontana, A., Castellano, M., {et~al.} 2017, ApJ, 847, 76,
  \dodoi{10.3847/1538-4357/aa8874}

\bibitem[{{Schaerer} {et~al.}(2018){Schaerer}, {Izotov}, {Nakajima}, {Worseck},
  {Chisholm}, {Verhamme}, {Thuan}, \& {de Barros}}]{Schaerer2018}
{Schaerer}, D., {Izotov}, Y.~I., {Nakajima}, K., {et~al.} 2018, \aap, 616, L14,
  \dodoi{10.1051/0004-6361/201833823}

\bibitem[{Shibuya {et~al.}(2015)Shibuya, Ouchi, \& Harikane}]{Shibuya2015}
Shibuya, T., Ouchi, M., \& Harikane, Y. 2015, ApJS, 219,
  \dodoi{10.1088/0067-0049/219/2/15}

\bibitem[{{Shimakawa} {et~al.}(2015){Shimakawa}, {Kodama}, {Steidel}, {Tadaki},
  {Tanaka}, {Strom}, {Hayashi}, {Koyama}, {Suzuki}, \&
  {Yamamoto}}]{Shimakawa2015}
{Shimakawa}, R., {Kodama}, T., {Steidel}, C.~C., {et~al.} 2015, \mnras, 451,
  1284, \dodoi{10.1093/mnras/stv915}

\bibitem[{{Song} {et~al.}(2016){Song}, {Finkelstein}, {Ashby}, {Grazian}, {Lu},
  {Papovich}, {Salmon}, {Somerville}, {Dickinson}, {Duncan}, {Faber}, {Fazio},
  {Ferguson}, {Fontana}, {Guo}, {Hathi}, {Lee}, {Merlin}, \&
  {Willner}}]{Song2016}
{Song}, M., {Finkelstein}, S.~L., {Ashby}, M. L.~N., {et~al.} 2016, \apj, 825,
  5, \dodoi{10.3847/0004-637X/825/1/5}

\bibitem[{{Speagle} {et~al.}(2014){Speagle}, {Steinhardt}, {Capak}, \&
  {Silverman}}]{Speagle2014}
{Speagle}, J.~S., {Steinhardt}, C.~L., {Capak}, P.~L., \& {Silverman}, J.~D.
  2014, \apjs, 214, 15, \dodoi{10.1088/0067-0049/214/2/15}

\bibitem[{Stefanon {et~al.}(2017)Stefanon, Bouwens, Labb{\'{e}}, Muzzin,
  Marchesini, Oesch, \& Gonzalez}]{Stefanon2017}
Stefanon, M., Bouwens, R.~J., Labb{\'{e}}, I., {et~al.} 2017, ApJ, 843, 36,
  \dodoi{10.3847/1538-4357/aa72d8}

\bibitem[{{Steidel} {et~al.}(2016){Steidel}, {Strom}, {Pettini}, {Rudie},
  {Reddy}, \& {Trainor}}]{Steidel2016}
{Steidel}, C.~C., {Strom}, A.~L., {Pettini}, M., {et~al.} 2016, \apj, 826, 159,
  \dodoi{10.3847/0004-637X/826/2/159}

\bibitem[{{Steidel} {et~al.}(2014){Steidel}, {Rudie}, {Strom}, {Pettini},
  {Reddy}, {Shapley}, {Trainor}, {Erb}, {Turner}, {Konidaris}, {Kulas}, {Mace},
  {Matthews}, \& {McLean}}]{Steidel2014}
{Steidel}, C.~C., {Rudie}, G.~C., {Strom}, A.~L., {et~al.} 2014, \apj, 795,
  165, \dodoi{10.1088/0004-637X/795/2/165}

\bibitem[{{Storey} \& {Zeippen}(2000)}]{Storey2000}
{Storey}, P.~J., \& {Zeippen}, C.~J. 2000, \mnras, 312, 813,
  \dodoi{10.1046/j.1365-8711.2000.03184.x}

\bibitem[{{Swinbank} {et~al.}(2019){Swinbank}, {Harrison}, {Tiley}, {Johnson},
  {Smail}, {Stott}, {Best}, {Bower}, {Bureau}, {Bunker}, {Cirasuolo}, {Jarvis},
  {Magdis}, {Sharples}, \& {Sobral}}]{Swinbank2019}
{Swinbank}, A.~M., {Harrison}, C.~M., {Tiley}, A.~L., {et~al.} 2019, \mnras,
  487, 381, \dodoi{10.1093/mnras/stz1275}

\bibitem[{{The Astropy Collaboration}(2018)}]{Astropy2018}
{The Astropy Collaboration}. 2018, {astropy v3.1: a core python package for
  astronomy}, 3.1, Zenodo,  Zenodo, \dodoi{10.5281/zenodo.4080996}

\bibitem[{{Treu} {et~al.}(2022){Treu}, {Roberts-Borsani}, {Bradac}, {Brammer},
  {Fontana}, {Henry}, {Mason}, {Morishita}, {Pentericci}, {Wang}, {Acebron},
  {Bagley}, {Bergamini}, {Belfiori}, {Bonchi}, {Boyett}, {Boutsia},
  {Calabr{\'o}}, {Caminha}, {Castellano}, {Dressler}, {Glazebrook}, {Grillo},
  {Jacobs}, {Jones}, {Kelly}, {Leethochawalit}, {Malkan}, {Marchesini},
  {Mascia}, {Mercurio}, {Merlin}, {Nanayakkara}, {Nonino}, {Paris},
  {Poggianti}, {Rosati}, {Santini}, {Scarlata}, {Shipley}, {Strait}, {Trenti},
  {Tubthong}, {Vanzella}, {Vulcani}, \& {Yang}}]{Treu2022}
{Treu}, T., {Roberts-Borsani}, G., {Bradac}, M., {et~al.} 2022, \apj, 935, 110,
  \dodoi{10.3847/1538-4357/ac8158}

\bibitem[{{Umeda} {et~al.}(2022){Umeda}, {Ouchi}, {Nakajima}, {Isobe},
  {Aoyama}, {Harikane}, {Ono}, \& {Matsumoto}}]{Umeda2022}
{Umeda}, H., {Ouchi}, M., {Nakajima}, K., {et~al.} 2022, \apj, 930, 37,
  \dodoi{10.3847/1538-4357/ac602d}

\end{thebibliography}


\end{document}